# ITCM: A Real Time Internet Traffic Classifier Monitor

Silas Santiago Lopes Pereira[1], José Everardo Bessa Maia[2] and Jorge Luiz de Castro e Silva[2]

[1]Federal Institute of Ceará, Aracati, Brazil
[2]Department of Statistics and Computing, State University of Ceará, Fortaleza, Brazil

## ABSTRACT

*The continual growth of high speed networks is a challenge for real-time network analysis systems. The real time traffic classification is an issue for corporations and ISPs (Internet Service Providers). This work presents the design and implementation of a real time flow-based network traffic classification system. The classifier monitor acts as a pipeline consisting of three modules: packet capture and pre-processing, flow reassembly, and classification with Machine Learning (ML). The modules are built as concurrent processes with well defined data interfaces between them so that any module can be improved and updated independently. In this pipeline, the flow reassembly function becomes the bottleneck of the performance. In this implementation, was used a efficient method of reassembly which results in a average delivery delay of 0.49 seconds, approximately. For the classification module, the performances of the K-Nearest Neighbor (KNN), C4.5 Decision Tree, Naive Bayes (NB), Flexible Naive Bayes (FNB) and AdaBoost Ensemble Learning Algorithm are compared in order to validate our approach.*

## KEYWORDS

*Traffic Classification System, Pipeline, Flow Reassembly, Machine Learning .*

## 1. INTRODUCTION

The Internet traffic is changing continuously and this contribute to difficult the characterization of network behaviour and structure. Massive games [1] and cloud and grid services increase every day their percentage participation in total network traffic. Traffic monitoring Systems generally make use of flow information. Examples are NetFlow [2] or IETF IPFIX [3], which defines a standard to exporting flow information by routers and switches. Such systems are widely used in network service providers and corporations to gain knowledge about critical business applications, analyze communication patterns prevalent in traffic, collect data for account, or detect anomalous traffic patterns [4]. A vital issue for corporations and ISPs (Internet Service Providers) is to identify traffic application types which are transmitted on their networks [5].

Pattern recognition and machine learning models have given significant attention to semi-supervised learning [6]. In network traffic areas, encryption and processing restrictions, protocol obfuscation and use of ephemeral ports make the task of construct classification models difficult. The large amount of Internet traffic flowing through networks makes the use of approaches that combine labelled and unlabeled data to construct accurate classifiers suitable.

There are a large number of papers in the traffic monitoring and traffic classification area. Most papers usually focus on either traffic flow reassembly or traffic classification and identification, but not on their combination. This paper describes the architecture of a real time Internet traffic classifier monitor for use in corporate networks. It also evaluates different machine learning methods for network traffic classification. The classifier monitor is based on concept of bidirectional flow. This means that the fundamental object to be classified in a determined pattern is the traffic flow, either complete or as subflow. A flow is defined by one or more packets between a host pair with the same quintuple: source and destination IP address, source and destination ports and protocol type (ICMP, TCP, UDP) [9].

The remainder of this paper is organized as follows. Section 2 overviews the related work about flow reassembly and traffic classification. In section3, we describe the design and implementation of the classifier monitor. Section 4 details the data collection used for evaluate the ITCM and describe how the experiments were performed. Section 5 presents and discusses the performance tests results. Section 6 ends with some conclusions and future work.

## 2. RELATED WORK

Statistical classification is based on collecting statistical information on properties of the traffic flow, and relies on the assumption that each category has a particular distribution of properties which represents it and can be used to identify it [10]. The statistical traffic classification using machine learning techniques have been widely explored in the recent years.

There are a limited number of tools available in the literature for traffic classification [7]. The NetAI tool is able to perform online and offline feature extraction, although not directly perform traffic classification. The FullStats is able to extract an extensive set of characteristics, but from a offline trace. The GTVS, which is a DPI (*Deep Packet Inspection*) based software, allows the labelling of traces in a semi-automated manner. The only two traffic classification tools which implement machine learning method are Tstat 2.0 and TIE. The Tstat makes use of the *packet size* and *interpacket time* features in a Bayesian framework for identification of Skype and obfuscated P2P file sharing. Although the tool has a limited number of applications, it can extract a large number of characteristics. The TIE software platform is available to the research community, and which allows the development of classification methods. The framework provides traffic capture and processing, feature extraction, and online classification. At the moment, a few number of features is available on TIE. Systems like Bro, which is able to collect flow statistics and perform payload-based classification at high speed rates, are limited when the traffic is encrypted [8]. Here, we briefly review some important approaches to stream reassembly (subsection 2.1) and traffic classification (subsection 2.2).

### 2.1. Stream Reassembly

In [11], the authors present an efficient TCP stream reassembly mechanism for real time network traffic processing at high speeds. The mechanism uses the recently-accessed-first principle to reduce the search cost of a connection for each packet arrival. Moreover, to improve the search process, the system keeps established and not established TCP connections in different structures. Experimental results based on network traffic captured in a typical gigabit gateway showed the proposed policy, in comparison of traditional one (RFC 793 [12]), was more efficient and could attend the real time property requisite of traffic analysis systems in gigabit networks.

In [13], the author presents a TCP stream reassembly mechanism designed and implemented to an network-based intrusion detection system. The system receives individual packets from network and performs signature detection from the payload.  The approach is described as follows: First, the system associates each received packet to its corresponding TCP connection, based on the quadruple composed of source IP address and port, and destination IP address and port. Then, the system checks the packet sequence number and determines if this packet is the next expected packet for the respective connection. If true, the packet is sent for signature detection.

In [14], to improve the forensic analysis, the authors present a TCP stream evaluation methodology which consists of estimate the TCP reassembly accuracy by the precise identification of potential errors at packet and stream levels hidden in this process. This approach can be used for derivation and computation of reassembly errors. In the proposed TCP reassembly model, an session counting algorithm is presented, which defines a flow as a set of TCP packets with same values to source IP and port, and destination IP and port, and a flow can have multiples sessions delimited by well defined phases of connection establishment and termination. From two traffic captures obtained with *Tcpdump* tool, the authors used a libpcap-based program [15] to read the packet traces and evaluate the reassembly and error verification approaches , which was experimentally validated with known analysis tools as *Tcptrace* and *Tcpflow*.

## 2.2. Traffic Classification

In [16], it is evaluated the effectiveness of machine learning techniques for the real time traffic classification problem using statistical attributes derived from first packets of each flow. The authors utilized port-base method for labelling flows into categories. Since the traffic traces are anonymized by privacy reasons, it precludes the inference of applications that generated the flows. Although this approach can consequently introduce incorrectly labelled flows, the authors argue that, for the studied ports, the non-labelled flow instances percentage is low and most of the traffic belongs to standard applications. The acquired results showed that the classification with decision trees had the highest accuracy and performance in comparison of other classifiers. In addition, subflow-based classifiers can reach high accuracy values while the computational complexity is reduced.

In [17], the machine learning approach applied to traffic classification using only transport-layer statistics is explored. This approach seeks to circumvent the problem that many network applications, such as P2P protocols, make use of dynamic port numbers and content encryption to avoid detection. This becomes inefficient the traditional approaches of port mapping and content analysis. The author evaluate the impact on the performance of data dimensionality, selected attributes, and machine learning used algorithms, which were, respectively, TAN, C4.5, NBTree, Random Forest and Distance Weighted KNN. The application of classification techniques and discriminant selection based on genetic algorithms together can dramatically reduce the learning and modelling time with little variation in the classification process accuracy.

Bar-Yanai et al. (2010) [10] proposes a hybrid combination of k-means and k-nearest neighbor geometrical classifiers. The proposed statistical classifier was integrated into a real-time embedded environment implemented on a SCE2020 Cisco platform and the algorithm was tested at a full line rate. The average accuracy rate of K-Means and k-nearest neighbor algorithm is of 83% and 99.1%, respectively, and the accuracy of the hybrid approach is very similar to the k-NN technique, with marginal accuracy loss. The time complexities of the hybrid solution are much lower than those of the k-NN algorithm, which has the better accuracy but higher complexity. The authors conclude that their proposed algorithm works well on platforms with limited memory, CPU resources and real-time requests. The technique works as follows: Initially, the training dataset is divided in N smaller datasets, each one corresponding to one of the known application categories presented in the training data. After this, each partition is grouped with the K-Means clustering method for a certain number of K clusters, so that each cluster is labelled naturally by the application protocol in which it was constructed. The second training stage consists in redistribution of the entire sample set of cluster centers defined in the previous stage. The online classification is performed by the execution of the KNN (k=1) classification method over the members of the closest cluster to an unseen flow.

Mallapradda et al. (2009) [6] propose a boosting framework for binary classification which combines advantages of graph-based approaches and ensemble methods. The strategy is to improve the performance of a given supervised learning algorithm using unlabeled data. The proposed algorithm is named *SemiBoost*, and is a general and efficient approach that allows the choice of a base classifier which is well-suited to a specific task. Like other boosting algorithms, the classification accuracy is improved iteratively, but SemiBoost selects unlabeled data along the iterations. The proposed algorithm combines similarity information with classifier predictions to obtain the most confident pseudo-labels. The authors used WEKA [18] software in the implementation of benchmark semi-supervised approaches. From the evaluation and comparison of the method with three state-of-art semi-supervised algorithms (TSVM, LDS and LavSVM) and in 16 different datasets indicate significant gain of the proposed approach, when evaluated with the base classifiers Decision Stump, J48 and SVM.

Szabó et al. (2011) [19] proposes a framework for traffic classification which uses machine learning techniques and only information from the packet headers. One component of the proposed framework is the combination of classification and clustering algorithms to make the identification system robust under different network conditions. The training and evaluation of classification system were performed with the traffic flow obtained from measurements in networks with different access technologies and different locations, in order to make the traffic characteristics varied. The authors found that these clustering and classification methods resulted with different performance results when used to identify traffic from unknown networks. They also verified that clustering algorithms have proved to be more robust with network parameter changes while classification methods can learn about a specific network more accurately. The authors present and evaluate two different combinations of classification and clustering approaches that result in accuracy increase when comparing to standalone cases. The first combination, named classification with clustering information, means that each training flow has its respective cluster number as a new attribute for supervised classification. The cluster numbers of training flows are obtained from the previous clustering of training data. Since the clustering information attribute may be neglected or considered with low-importance by a supervised technique, this approach cannot always improve the global accuracy. The second approach, named model refinement with per cluster based classification, initially applies unsupervised learning to generate clusters. A separate classification model is then built for the set of flows of each cluster. In the evaluation phase at an unseen flow, the unsupervised method results with the number for the most similar cluster for which the associated model is used to evaluate the flow. This approach always considers the clustering results with high importance and the supervised techniques can build simple models since each group contains a limited number of flow types. This implies that the impact of over-fitting with the classification model is reduced. The per cluster based classification scheme outperforms the standalone supervised and unsupervised methods. The proposed methods obtained 93% and 75% of TP ratio for the evaluation on the same network and the cross-checks on other networks, respectively.

Erman et al. (2007) [8] proposes and evaluates a semi-supervised methodology for statistical traffic classification that is able to accommodate known and unknown applications. The authors mention three main advantages to their proposed semi-supervised method: Initially, fast and accurate classifiers can be designed since a training dataset that consists with a small number of labelled flows and a large number of unlabeled flows. Second, the approach is robust and can handle both previously unknown applications and new patterns of existing applications. Moreover, network operators can insert unlabeled flows to improve the classifier performance, allowing the iterative development. Finally, the proposed approach can be integrated with solutions that collect flow statistics. The semi-supervised model combines supervised and unsupervised methods in two steps: The approach first uses the K-Means clustering technique to partition a training dataset composed of a few labelled flows and abundant unlabeled flows. The second step utilizes the available labelled flows to build a cluster with application mapping for which clusters without labelled flows remain unmapped, which corresponds to flows that possibly do not belong to any known application. The authors found that the proposed model was able to identify a variety of different applications with a high rate of accuracy, such as Web, P2P, FTP and E-mail. The flow and byte accuracy was above 98% and 93%, respectively. Furthermore, the authors noted that datasets with large number of flows consistently achieve a high classification of accuracy. The authors verify that despite labelling tools, labelling a large dataset can be expensive and difficult. In practice, labelling only a fraction of the training flows is sufficient to obtain high levels of accuracy.

## 3. THE CLASSIFIER MONITOR

This section details the design and operation of our classifier monitor, and concludes with the presentation of the modules which compose the system.

### 3.1. Architecture

The monitor works as a three-stage pipeline, with a collect and preprocessing module, a flow reassembly module, and an attribute extraction and classification module. For the purpose of pipeline, the time is divided in intervals of 30s. This value was chosen arbitrarily. Once the monitor starts, three parallel processes are in execution on each interval: the packet capture, the flow reassembly of the previous interval packet capture, and the flow classification for the collection occurred in two delay intervals. Another parallel process is responsible for closing old connections periodically, in order to reduce the use of memory and processing during reassembly process. This approach allows the classifier monitor reach a response time of $(30 + \alpha)$ seconds, where $\alpha$ is the necessary time to performs the reassembly of the captured data in a given interval. In summary, the monitor works with a *quantum* of 30s of traffic and with an average delay of $\alpha$ seconds in the flow reassembly, feature extraction, and classification. The average found value achieved in the current implementation was $\alpha = 0.49$ seconds.

The Figure 1 exhibits the capturing and processing environment of the monitoring and classification system. Basically, we assume that the traffic is mirrored by a network border router to a network interface monitored by the system. The system periodically performs the processing and categorization of captured data, and presents the obtained information from monitoring and classification process.

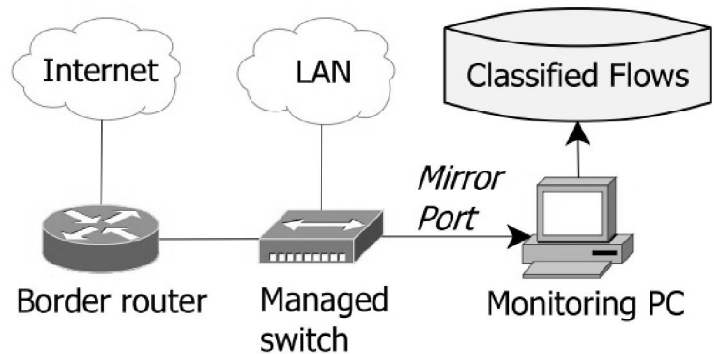

Figure 1. Traffic Capture Environment

The Figure 2 exhibits the layered structure of implemented classifier monitor, whose tasks modules are online traffic collection from a network point, pre-processing for flow reassembly, extraction and selection of statistical attributes, flow labelling since payload analysis or port-based method (only during training step), training with a supervised machine learning technique and classification, using the ML model built from training data. The classifier monitor performs the packet traffic capture continuously. In the training phase, the captured packets are sent to a reassembly process, which associate each packet to its respective flow. A parallel process extracts statistical information from the packet headers, selects the most relevant attributes using an attribute selection algorithm, and labels the flows with well-known ports method [20]. The traffic flows, which are disposed in a spatial representation (each flow is an instance with a set of characteristics), are used to training a selected supervised classification method. In the evaluation phase, the unlabeled flows obtained with the collection, reassembly, and attribute extraction are finally evaluated by the classifier.

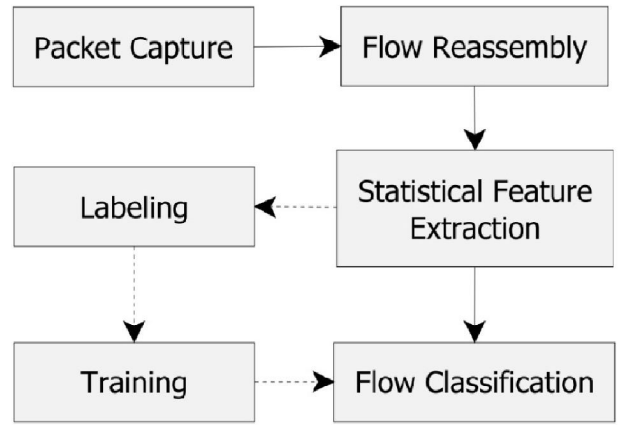

Figure 2. Block Diagram of Classifier Monitor

Since the modules are implemented as concurrent processes, the data interfaces between each module allows that each one can be improved and updated independently. To the packet capture and pre-processing, the input and output are respectively the TCP packet and a data structure with only the necessary packet information to the monitor (e.g. bytes, timestamp, flags, no payload). For the flow reassembly module, the input and output are respectively a set of pre-processed packets and the data structure which represents the reconstructed flow. For the classification module, the features vector is the input, and the output is the class.

The monitor was developed in C#.Net programming language using Visual Studio Integrated Development Environment. The online packet capture is performed based on sequential reading and processing of each packet contained from a network interface. The monitor also adopts a given timeout for packet capture and presentation of results. The reason for implementation of a flow reassembly algorithm, despite given the existence of several tools and libraries to reach this goal, as libNIDS [21], TcpTrace [22], and WireShark [23], is the possibility to evaluate different approaches for subflow classification, as mentioned in [24]. Moreover, the evaluation of approaches for real time TCP stream reassembly becomes possible, as in [25] and [26], which are fundamental in the development of a high-speed traffic classification system.

## 3.2. Packet capture and Pre-processing

The traffic capture at packet level, following by data processing and visualization, is a common demand on traffic volume monitoring or traffic supervision tasks. Therefore, it is essential that packet capture process must capture completed packets so that the stream is reassembled correctly.

For the task of packet capture, we use "TCP Session Reconstruct Tool" [27], which is a C# utility for packet capture and reconstruction of complete and incomplete TCP sessions. This tool is available with CPOL license, and is based on libnids library [21] and Wireshark. It uses a reconstruction TCP algorithm named TcpRecon. TcpRecon reconstructs a bidirectional flow, holds each flow in a dictionary structure and recovers its payload. We reuse this software by replacing the TcpRecon policy by our proposed reassembly scheme. The TCP flow Reassembly is explained in the next subsection.

## 3.3. Flow Reassembly

A reassembly function associates a TCP packet with its respective stream. The purpose of this function is to recover the initial state, emitted by sender, from the captured TCP packets [14]. It is essential that reassembly, which is applicable to a diversity of network traffic analysis systems, such as intrusion detection and prevention, content inspection and network forensics, can be performed as fast as possible to handle high traffic rates, especially in high-speed networks [11]. Although the RFC 973 [14] presents a standard specification of TCP protocol (the TCP protocol is specified in a large number of RFCs), there are different implementations, which make TCP reassembly a difficult task. Different reassembly tools hold their own stream concept specifications. For example, the Tcpflow tool binds a tuple with a stream, while the Tcptrace tool associates a session with a given stream. The Tcptrace and Tcpflow tools group the data sent in each direction in different streams. In a stream generated by the Wireshark tool, data from sender and receiver are grouped in the same stream.

### 3.3.1. Recently-accessed-first Principle

In a high-speed network, there may be hundreds of thousands of simultaneous connections, so that the reassembly system, which maintains a structure for storing connection records, searches in this structure the corresponding record for each collected packet. The search becomes expensive and needs to be optimized as the number of connections increases [11]. The idea of recently-accessed-first principle is to bring the most recently accessed connection records to top of connection record list.

Since in a TCP connection the data transfer obeys the TCP/IP specification, the locality principle applied to the arrival of packets on network is based on assumptions that, given a packet and its subsequent, it is presumable both belong to the same connection, and considering the packet arrival of a certain connection, the next packet of this connection will come soon [11]. From the described principle about TCP packet arrival, the recently-access first principle is adopted in search of connection record set. For every successful search, the accessed record is moved to top of the set, so that after a sequence of access, the most frequently accessed nodes will be in the beginning of set. Thus, they can be accessed more quickly and the search efficiency is improved [11].

This principle, although efficient for clients, present bad performance on server or links with a lot of traffic flows. This is because the locality principle is lost in these scenarios [28]. However, a concurrent process in our monitor finishes old flows periodically in order to circumvent this issue. This work uses the same concept of TCP stream presented in [14], and the recently-access-first principle presented in [11] to optimize the reassembly step of our software-based solution. Differently of [11], which uses two hash tables for connection management, we use a simple list to store connections. The adopted reassembly policy was based on mechanism proposed in [13] for TCP session reassembly. The applied reassembly approach was validated experimentally with the *Tcptrace* and *Tcpflow* tools. The reassembly approach is described in algorithm on Figure 3 and works as follows: for each TCP packet received, the system searches the corresponding connection in the connection record list. If the record is valid, the package is inserted into this one. If is an invalid record and the packet contains a SYN flag, a new connection is created for the packet. If the record is invalid and the packet does not contain the SYN flag, the packet is dropped. If the packet containing the FIN or RST flags, the connection is terminated.

$Packet \leftarrow GetPacket()$
$quintuple \leftarrow Packet.Quintuple$
**if** Packet.SYN **then**
    $C \leftarrow GetNotEstablishedConn(quintuple)$
    **if** $C$ is invalid **then**
        $conn \leftarrow NewConnection(Packet)$
        $InsertNotEstablishedConn(conn)$
    **else**
        $InsertPacket(C, Packet)$
    **end if**
**else**
    $C \leftarrow GetEstablishedConn(quintuple)$
    **if** $C$ is invalid **then**
        $C \leftarrow GetNotEstablishedConn(quintuple)$
        **if** $C$ is invalid **then**
            $DropPacket(Packet)$
        **else**
            $InsertInEstablishedConn(C)$
            $RemoveFromNotEstablishedConn(C)$
            $InsertPacket(C, Packet)$
            **if** Packet.FIN or Packet.RST **then**
                ConcludeConn(C, Packet)
            **end if**
        **end if**
    **else**
        $InsertPacket(C, Packet)$
        **if** Packet.FIN or Packet.RST **then**
            ConcludeConn(C, Packet)
        **end if**
    **end if**
**end if**

Figure 3. TCP Flow Reassembly Algorithm

### 3.4. Flow Labelling

Labelling is a necessary step for training and subsequent evaluation of classifiers. Although the utilization of port-based method [20] to traffic flow labelling can introduce errors due its increasing ineffectiveness since flows can be incorrectly labelled, the existence of some inaccurate values in data sets is a common machine learning problem, and a good ML scheme must have the capability to deal with this situation [16]. Although this labelling method was implemented in the first prototyping of ITCM, other sophisticated labelling techniques can be subsequently incorporated.

## 3.5. Traffic Classification

Real time internet traffic classification enables the solution of hard network management problems by Internet service providers and their equipment suppliers. Network operators, especially in high-speed networks, need to know about the current traffic to quickly respond to support diverse business goals [29].

The presented approach uses real traces in the evaluation of Naive Bayes, Kernel Naive Bayes, C4.5 Decision Tree and K-Nearest Neighbor methods for Internet traffic classification using statistical information derived from packet headers. The Weka [18] (Waikato Environment for Knowledge Analysis) is an Open Source tool implemented in JAVA, and contains a collection of machine learning algorithms for data mining problems. The ITCM utilizes the Weka libraries for the training and evaluation of the machine learning methods. We utilize the IKVM [30] software to enable Java and .NET interoperability. This tool allows to generate the Weka *dlls*, which were imported to C# code. Thus, it is possible to use the weka classifiers in the classification module of our classifier monitor.

### 3.5.1. K-Nearest Neighbor

Among the various supervised statistical methods for pattern recognition, the Nearest Neighbor (NN) technique is the one that achieves better results, without priori assumptions requirement about the training samples distributions [31]. The algorithm assumes all instances correspond to points on an n-dimensional space $\mathbb{R}^n$. A new instance $X = \{x_1, x_2, \ldots, x_n\}$, in which $x_1, x_2, \ldots, x_n$ are the corresponding attributes, is classified by computation of its Euclidean distance to the training instances, and then categorized with the label of the nearest training instance [17].

The KNN *(K-Nearest Neighbors)* classifier extends this idea through the selection of the $k$ nearest neighbors and the classification of each new instance with the most common class among them [31]. The Euclidean distance between two instances $X$ and $Y$ is defined by the following expression, where $x_k$ and $y_k$ denote, respectively, the values for the k-th attribute of instances $X$ and $Y$:

$$d(X, Y) = \sqrt{\sum_{k}^{n} (x_k - y_k)^2}$$

### 3.5.2. Naive Bayes

The NB classifier is a simple technique that can be applied to the Internet traffic classification problem [32]. Consider a random variable $C$ that denotes the class on an instance, and a random variable vector $X$ which represents the observed values of attributes. Furthermore, assume $c$ a label of a determined class and $x$ an attribute values vector. Consider a test instance $x$ to be classified. The most probable class will be the one of most high value to $P(C = c \mid X = x)$, which is the probability of occurrence of class $c$ class given the instance $x$. The following expression presents the Bayes rule, applied to calculate this probability, where $X = x$ corresponds to $X_1 = x_1 \wedge X_2 = x_2 \wedge \ldots X\_k = x_k$ event, and $P(C = c)$ represents the priori probability of $c$, which is the probability of obtain the class $c$ without consider the training data:

$$p(C = c | X = x) = \frac{p(C = c) p(X = x | C = c)}{p(X = x)}$$

A common assumption, which is not inherent in Naive Bayesian approach but still often used, is that for each class, the numerical attributes values are normally distributed. According to [33], despite this assumption does not reflects the Internet traffic reality, such approach outperforms some more complex models.

### 3.5.3. Kernel Naive Bayes

The KNB learning technique is a generalization of the NB and algorithmically similar to this one in every aspects, except in computation of distribution function $P(X = x | C = c)$ for continuous attributes, which can be replaced by a variety of non-parametric estimation methods, among them, the kernel density estimation, which, as the name suggests, uses kernel estimation methods instead of simple Gaussian approach [34]. The distribution function is shown by the following expression, where $n$ represents the number of training instances that belongs to class $c$, and $u_i = x_i$:

$$p(X = x | C = c) = \frac{1}{n} \sum_{i}^{n} g(x; \mu_i, \sigma_c)$$

The intention of KNB method is that the kernel estimation allow the technique has a good performance in domain that violate the normality assumption. According to [9], the simple assumption about the normality of the discriminators is inaccurate and eminent problems arise when the actual distribution is multimodal, and this situation may indicate that the considered class is too large or other distribution must be used for the data analysis.

### 3.5.4. C4.5 Decision Tree

The C4.5 method, which is used to generate Univariate decision tree [35], is a statistical classifier because its decision trees can be used in classification [36]. The classifier was created by Ross Quinlan [37] and is an extension of ID3 (Iterative Dichotomizer 3) algorithm, which builds simple decision trees. C4.5 makes use of information entropy to build decision tree in the same way as the ID3.

The C4.5 makes various decisions based on data takes into account all available input features [38]. The algorithm recursively chooses the feature with highest normalized information gain at each node of the tree. The chosen feature splits the data into subsets enriched at one class or the other [36]. The information gain indicates how well a decision will separate the output class from most [38]. We utilized the J48 learning method, which is an open source Java implementation of the C4.5 algorithm.

### 3.5.5. AdaBoost Ensemble Learning Algorithm

AdaBoost is an ensemble learning algorithm which alters weak learning classifiers by assigning weights with the overall goal to correctly classify the instances. Generic Boosting manipulate de training data in order to generate distinct classifiers and improves the classification accuracy iteratively. At each iteration, the weights are ajusted so that the weights of the misclassified examples are increased and those of correctly classified examples are decreased. All instances are used in each iteration. Each classifier generated at each round of AdaBoost has an associated weight, which means that a classifier with high degree of correctly identified instances will have more weight in the total ensemble. The final decision of the ensemble is the weighted-majority voting of the hypothesis generated of each generated classifier [38]. See [39] for more details.

# 4. METHODOLOGY

This section details the design and operation of our classifier monitor, and concludes with the presentation of the modules which compose the system.

## 4.1. Data Collection and Experiments

The performance of packet capture and reassembly modules was evaluated for capacity verification, under variable load conditions. The classifier monitor was performed with a Core i5 computer with CPU 2.30 GHz and 4GB of memory. The trace-driven simulation allows flexibility in the evaluation of distinct classifiers and reassembly approaches. This is because the different executions of our system for the same packets trace always generate the same flows set. Without this determinism, it would be extremely difficult to reproduce the same results of an online packet capture, given the possibility of delay and packet loss, for example. For confidently evaluate the online monitor, we used traffic traces collected in a host connected to the broadband Ethernet 100Mbps. Each flow in a reassembly process is configured with a 60 seconds timeout, in order to avoid the storage of old or idle connections, which consume memory and processing resources unnecessarily. This means that TCP flows whose duration is greater than this value are finished by the collector process periodically. We compare the time complexities and the number of reconstructed flows of our flow reassembly module and the external tools Tcptrace, TcpFlow, TcpRecon and Wireshark. Using Weka resources, we use 10-fold cross validation for accuracy evaluation of the aforementioned classification models.

The used traffic traces characteristics, referred as $T_1$ and $T_2$, are presented in Tables I and II. They reflect the communication between a host at State University of Ceará (UECE) and the Internet.

Table 1. Characteristics of the Trace $T_1$

| Characteristic | Description |
|---|---|
| Packets Number | 614282 |
| Capture Size | 565.62MB |
| Capture Duration | 3516.79s |
| Average packet Size | 920.78 Bytes |
| Average capture Rate | 1.28 Mbps |

Table 2. Characteristics of the Trace $T_2$

| Characteristic | Description |
|---|---|
| Packets Number | 1579921 |
| Capture Size | 1.88GB |
| Capture Duration | 1355.83s |
| Average packet Size | 1195.16 Bytes |
| Average capture Rate | 11.14 Mbps |

Table 3 presents the summary of the identified applications flows in the current traces, which were: Www (World Wide Web), Https (Http protocol over TLS/SSL), Ftp (File Transfer Protocol), Xvttp (Xvttp Protocol) and Isakmp (Isakmp Protocol). The most representative categories in $T_1$ traffic traces are Www and Ftp applications. In $T_2$ traffic trace, the Https and Isakmp application have a greater number of flow instances. In our study, the classification and training steps are performed at the end of packet capture simulation and reassembly. This methodology is necessary to evaluate the modules of our classifier monitor.

Table 3.  Category of Applications

| Category | Description | $T_1$ | $T_2$ |
|----------|-------------|-------|-------|
| Www-http | World Wide Web HTTP | 1022 | 337 |
| HTTPs | Http Protocol over TLS/SSL | 139 | 27 |
| FTP | File Transfer Protocol | 1808 | - |
| Domain | Domain Name Server | - | 1 |
| Total | | 2969 | 365 |

## 4.2. Statistical Features

In order to evaluate the classification process, we considerate the following features: elapsed time between the first and last packets, number of packets, number of bytes, the number of all packets with at least a byte of TCP data payload, the number of all packets seen with the PUSH bit set in the TCP header, and the median and the variance of the number of bytes in IP packet. Since each attribute is computed for both directions of flow (uplink and downlink), each flow instance has 14 statistical discriminators plus the class label. There was not a proper selection of attributes in this study. We chose some of the most often found attributes in previous published work [9] which could be calculated from the data contained in the header of packets without examining their payload.

## 5. RESULTS AND DISCUSSION

Table 4 presents some performance metrics of our classifier monitor for the used traces. We observed that the highest achieved throughput for the packet capture and reassembly modules was 3.95 flows per second (fps) for $T_1$ traffic trace. This means that this is the number of traffic flows are delivered by reassembly process at every second. Although this is a low value, since the mean packet capture throughput of $T_1$ is only 1.28 Mbps, the reassembly process achieves a throughput of 24997.25 fps at one of packet capture intervals. The average packet capture and reassembly rate, expressed by Mbits/($T_{CO}+T_{RE}$), was 1014.73 Mbps, where $T_{CO}$ and $T_{RE}$ are the total duration times of packet capture and reassembly, respectively Similarly, the same performance metrics are presented for $T_2$ traffic trace. We can observe no bottlenecks in the reassembly process, which could support the considerate traffic. Our software-based monitor is effective to work in real time for a corporate network, for example. The average delivery delay $\alpha$ for $T_1$ and $T_2$ traffic traces is 0.49s and 7.50s, respectively. This means that this is the average reassembly duration for ever quantum of 30s. We can observe that delivery delay varies greatly between the two compared traces. Although they are from the same packet capture point, the two trace files are essentially different. The throughput collection of the $T_2$ is much larger than $T_1$. Furthermore, the traffic load in $T_2$ is larger than $T_1$. This means that there are less traffic to process in $T_1$, and consequently, its delivery delay is lower than the other trace.

Table 4.  Performance of Monitoring and Reassembly Processes

| Metric | $T_1$ | $T_2$ |
|--------|-------|-------|
| TCP Connections Number | 2969 | 365 |
| Max Capture & Reassembly Throughput | 3.95 fps | 2.89 fps |
| Max Reassembly Throughput | 24997.25 fps | 128.21 fps |
| Average Capture & Reassembly Rate | 1014.73 Mbps | 635.34 Mbps |
| Average Delivery Delay | 0.49s | 7.50s |
| Total Monitor Time | 75.67s | 310.84s |

In Table 5, our system is compared with the Tcptrace, TcpFlow, TcpRecon and Wireshark tools. The TcpRecon was modified to use a flow timeout of 60 seconds. We can observe that the number of flows is not the same between the tools, because the divergence of the used traffic flow concept, as explained previously. We can observe that our adopted reassembly approach execution time is lower than the other tools. Our reassembly scheme was implemented in TCP session Reconstruction Tool, replacing the TcpRecon default policy. In summary, the difference between these two policies is the adopted recently-accessed-first principle and the use of different data structures to hold established and not established TCP connections.

Table 5. Comparison with External Tools

| Approach/Tool | Flows Number | Reassembly Time |
|---|---|---|
| Proposed Approach | 2969 | 75.67s |
| TcpFlow | 5894 | 118.87s |
| Tcptrace | 3044 | 612.95s |
| Wireshark | 3036 | 182.69s |
| TcpRecon | 3036 | 96.97s |

Since TcpRecon and our proposed scheme are written in same language and uses the same packet capture libraries, we also compare the performance of these two policies one of each other. The confidence interval estimation of an event population will have greater reliability if the event is executed at least 30 times [40]. We executed and measured the elapsed times of the aforementioned TCP reassembly policies. The policies were evaluated over the already presented datasets. We computed the average execution time and confidence level for each TCP policy. We consider a high confidence level of 95%. The resulting confidence levels for TcpRecon and our adopted reassembly scheme are presented in Table 6. For the $T_1$ traffic trace, our scheme obtain a time complexity advantage of 20.31 seconds. For the $T_2$ traffic trace, there is also a reduction of 9.39 seconds with our approach.

Table 6. Performance Comparison of Reassembly Policies

| Traffic Trace | TcpRecon | Proposed Policy |
|---|---|---|
| $T_1$ | $91.85 \pm 5.61$ seconds | $71.54 \pm 3.57$ seconds |
| $T_2$ | $380.36 \pm 14.82$ seconds | $370.97 \pm 7.72$ seconds |

The Table 7 presents the main results about the classification process. We can observe that C4.5 Decision Tree was able to categorize on average 87.40% and 89.86% of the traffic correctly for the two traffic traces. The AdaBoost ensemble algorithm, using the DecisionStump classifier, was able to categorize on average 78.17% and 89.58% of the traffic correctly. The KNN technique, with k=10, was able to categorize on average 86.25% and 91.50% of the traffic correctly, against 72.48% and 80.00% for NB classifier. The duration of classification phase was a few seconds, and the results aim to validate the previous phases of the classifier monitor.

Table 7. Global Accuracy per Trace

| Classifier | $T_1$ | $T_2$ |
|---|---|---|
| C4.5 Decision Tree | 87.40% | 89.86% |
| AdaBoost (DecisionStump) | 78.17% | 89.58% |
| K-Nearest Neighbor | 86.25% | 91.50% |
| Naive Bayes | 72.48% | 80.00% |
| Flexible Naive Bayes | 64.09% | 88.76% |

## 6. CONCLUSIONS

This paper presented the architecture, implementation, and performance of an Internet traffic classifier monitor. The monitor is composed of three modules which were implemented as concurrent processes: capture and pre-processing, flow reassembly, and classification. For the $T_1$ traffic trace, the throughput reassembly module of the current implementation is 24997.25 flows per second. The average delivery delay is 0.49 seconds. For the classification module, the C4.5 algorithm outperforms KNN and AdaBoost classifiers with average accuracy of 87.40% and 89.86% against 72.48% and 80% for the KNN and AdaBoost methods, respectively.

Future directions for this research includes to incorporate subflow based classification in ITCM to reduce response time. Second, we aim to verify the performance impact of our classifier monitor at gigabit links, which are becoming increasingly common at computer networks. Finally, we could also prototype ITCM with NetFPGA hardware, since the implementation of network systems in hardware is essential for any real time application, particularly in gigabit networks [41].

## Authors

**Silas Santiago Lopes Pereira** received his B.S degree in Computer Science from State University of Ceará, Brazil, in 2010. Master degree in Computer Science from State University of Ceará, Brazil, in 2013. He is now a professor at Federal Institute of Ceará. His interests include machine learning applications and computer networks.

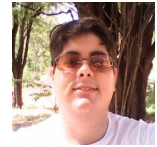

**José Everardo Bessa Maia** received his B.S. degree from Federal University of Ceará, Brazil, in 1980, M. S degree from State University of Campinas, Brazil, 1989, both in Electrical Engineering. Ph.D from the Federal University of Ceará, Brazil, 2011. He is a professor in the Statistics and Computer Science Department at State University of Ceará and University of Fortaleza.

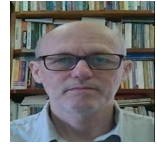

**Jorge Luiz de Castro e Silva** received his postdoctoral research in Computer Science from the State University of Campinas, Brazil, 2011. Ph.D in Computer Science from the Federal University of Pernambuco, Brazil, 2004. Master degree in Computer Science from the Federal University of Ceará, Brazil, 1997. Received his B. S. degree in Data Processing Technology from Federal University of Ceará, Brazil, 1978, and B.S degree in Business Administration from the State University of Ceará, Brazil, 1981. He is a professor at the State University of Ceará and Teacher / Advisor of the Master Program in Computer Science.

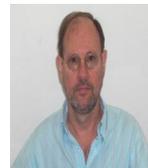